\documentclass[sigplan,10pt]{acmart}

\renewcommand\footnotetextcopyrightpermission[1]{}
\setcopyright{none}

\AtBeginDocument{
  }

\usepackage{amsmath}
\usepackage{listings}
\usepackage{xcolor}
\usepackage{multirow}
\usepackage{enumitem}
\usepackage{xspace}
\usepackage{threeparttable}
\usepackage[linesnumbered,ruled,vlined]{algorithm2e}
\usepackage{placeins}
\usepackage{dblfloatfix}
\usepackage{float}
\usepackage{mathtools}
\usepackage{booktabs}
\usepackage{pifont}
\usepackage{bussproofs}
\usepackage{graphics}
\usepackage{natbib}

\usepackage{tikz}
\usetikzlibrary{arrows.meta, positioning, fit, backgrounds, shapes.geometric, calc}
\usepackage{colortbl}
\usepackage[most]{tcolorbox}
\newtcolorbox[auto counter, number freestyle={\noexpand\arabic{\tcbcounter}}]{definedbox}[2][]{enhanced,
    unbreakable,
    colback=white,
    colframe=black!75!white,
    title=Prompt~\thetcbcounter: #2,
    #1
}

\newcommand{\tl}[1]{}
\newcommand{\wgy}[1]{}
\newcommand{\yy}[1]{}
\newcommand{\cwn}[1]{}
\newcommand{\hfwei}[1]{}
\newcommand{\cwnres}[1]{}

\newcommand{\tool}{{\textsf{IC3Syn}}\xspace}

\newcommand{\hide}[1]{}
\newcommand{\greycode}[1]{\texttt{#1}}

\definecolor{tla-keyword}{RGB}{0,80,160}
\definecolor{tla-comment}{RGB}{80,120,80}
\definecolor{tla-string}{RGB}{160,32,32}
\definecolor{tla-linenum}{gray}{0.50}
\definecolor{tla-bg}{RGB}{255,255,255}
\definecolor{tla-frame}{RGB}{180,180,200}

\lstdefinelanguage{TLAPlus}{
  morekeywords={VARIABLES,VARIABLE,CONSTANTS,CONSTANT,EXTENDS,MODULE,
    INSTANCE,LOCAL,LET,IN,IF,THEN,ELSE,CASE,OTHER,CHOOSE,EXCEPT,
    DOMAIN,ENABLED,UNCHANGED,SUBSET,UNION,BOOLEAN,TRUE,FALSE,
    THEOREM,ASSUME,PROOF,BY,DEF,QED,SUFFICES,PICK,HAVE,TAKE,WITNESS,
    None},
  sensitive=true,
  morestring=[b]",
}

\lstdefinestyle{tla}{
  language=TLAPlus,
  basicstyle=\small\ttfamily,
  keywordstyle=\bfseries,
  commentstyle=\itshape,
  stringstyle={},
  numberstyle=\scriptsize\color{tla-linenum},
  numbers=left,
  numbersep=8pt,
  xleftmargin=14pt,
  framexleftmargin=14pt,
  backgroundcolor=\color{white},
  frame=none,
  aboveskip=0pt,
  belowskip=0pt,
  columns=fullflexible,
  keepspaces=true,
  showstringspaces=false,
  escapeinside={(*}{*)},
  literate=
    {/\\}{{{\textbf{/\textbackslash}}}}2
    {\\/}{{{\textbf{\textbackslash/}}}}2
    {=>}{{{\textbf{=>}}}}2
    {==}{{{\textbf{==}}}}2
    {\\in}{{{$\in$}}}2
    {\\E}{{{$\exists$}}}2
    {\\A}{{{$\forall$}}}2
    {|->}{{{$\mapsto$}}}2
    {~=}{{{$\neq$}}}2,
}

\begin{document}

\title{Synthesizing Inductive Invariants for Distributed Protocols via IC3 and Large Language Models}

\author{Weining Cao}
\orcid{0009-0006-2433-4920}
\affiliation{\institution{Nanjing University}
  \city{Nanjing}
  \country{China}
}
\email{weiningcao@smail.nju.edu.cn}

\author{Guangyuan Wu}
\orcid{0009-0003-4689-4097}
\affiliation{\institution{Nanjing University}
  \city{Nanjing}
  \country{China}
}
\email{guangyuanwu@smail.nju.edu.cn}

\author{Yuan Yao}
\orcid{0000-0002-6913-6542}
\affiliation{\institution{Nanjing University}
  \city{Nanjing}
  \country{China}
}
\email{y.yao@nju.edu.cn}

\author{Hengfeng Wei}
\orcid{0000-0002-0427-9710}
\affiliation{\institution{Hunan University}
  \city{Changsha}
  \country{China}
}
\email{hfwei@hnu.edu.cn}

\author{Taolue Chen}
\orcid{0000-0002-5993-1665}
\affiliation{\institution{Birkbeck, University of London}
  \city{London}
  \country{United Kingdom}
}
\email{t.chen@bbk.ac.uk}

\author{Xiaoxing Ma}
\orcid{0000-0001-7970-1384}
\affiliation{\institution{Nanjing University}
  \city{Nanjing}
  \country{China}
}
\email{xxm@nju.edu.cn}

\renewcommand{\shortauthors}{Cao et al.}


\begin{abstract}

  Distributed protocols are notoriously difficult to verify correctly. Proving safety typically requires inductive invariants that both imply the desired property and are preserved by every protocol transition; yet inferring such invariants remains a major bottleneck: existing approaches either restrict the protocol models to a decidable fragment of first-order logic or demand expert-crafted templates.

  We present \tool, a neuro-symbolic framework that synthesizes inductive invariants by executing an IC3-style process over TLA+ states with the assistance of Large Language Models (LLMs). At large, \tool combines a symbolic IC3 controller, which decomposes invariant synthesis into focused blocking tasks and an LLM which provides protocol-level reasoning that IC3 alone lacks for TLA+ specifications. This integration enables a disciplined yet flexible search for invariants without imposing logical restrictions or requiring manual templates.

  We evaluate \tool on 29 distributed protocols spanning consensus, reconfiguration and client–server systems, and compare it against Endive, IC3PO, SWISS and DistAI. \tool discovers candidate invariants for all 29 protocols, including MongoLoglessDynamicRaft (MLDR), an industrial-scale Raft-based reconfiguration protocol for which none of the compared tools reports a solution, as well as one complex Paxos variant. In each case, the invariants synthesized on finite instances are shown in TLAPS to be inductive for the full unbounded protocol, thereby establishing safety.

\end{abstract}

\maketitle
\pagestyle{plain}

\section{Introduction}

Distributed protocols lie at the core of modern infrastructure, but establishing their correctness is notoriously challenging. Safety bugs are often rare, state-dependent, and expensive to reproduce, making even extensive testing unreliable. As a result, formal verification is widely adopted: protocols are specified in languages such as TLA+~\cite{lamport2002specifying}, and safety can be established, for instance, by model checking of finite instances with TLC, SMT-based symbolic analysis with Apalache, or deductive proof in TLAPS~\cite{chaudhuri2008tlaps}. These approaches typically rely on \emph{inductive invariants}: state predicates that (1) hold in every initial state, (2) are preserved by every protocol transition, and (3) imply the target safety property. In practice, however, the main difficulty lies not in checking a given candidate, but in automatically \emph{inferring} one.

Existing automated approaches to inductive invariant synthesis generally involve a trade-off. Some restrict protocol models to decidable fragments of first-order logic (e.g., EPR~\cite{padon2017paxos}) to make invariant search tractable. These restrictions, however, exclude common TLA+ modeling patterns and may distort natural protocol structure. Other approaches, such as Endive~\cite{schultz2022kvy}, operate directly on TLA+ specifications but require manually curated predicate pools, thereby shifting much of the synthesis burden back to the users. In practice, discovering inductive invariants for distributed protocols remains one of the most labor-intensive steps in formal verification, often requiring weeks to months of expert effort~\cite{hawblitzel2015ironfleet,wilcox2015verdi}.

State-of-the-art LLMs demonstrate protocol-level reasoning ability to assist in this process. However, directly asking an LLM to produce a complete inductive invariant remains ineffective: the clauses must satisfy initiation, consecution, and safety implication, and any gap in the conjunction invalidates the entire invariant. What is therefore needed is a \emph{principled decomposition} that reduces this global synthesis task into a sequence of localized, focused subproblems.

IC3 (aka Property-Directed Reachability, or PDR~\cite{bradley2011sat}) is a generic verification method for proving safety properties of transition systems. One key observation is that it provides precisely the kind of decomposition needed to synthesize inductive invariants for distributed protocols. Rather than requiring all protocol-level facts at once, IC3 constructs invariants incrementally: it maintains an explicit sequence of frame over-approximations and progresses by identifying bad states (i.e., states in the frontier frame that can transition to a safety violation). For each such state, IC3 performs a backward search through predecessor states in earlier frames; only when the search reaches a \emph{predecessor-free} state does it learn a \emph{blocking clause}, which is then validated and propagated forward. Each blocking step is local and targeted.

\medskip
\noindent\emph{Our work.} We propose \tool, a neuro-symbolic framework that executes an IC3-style process over TLA+ states to synthesize inductive invariants with the assistance of LLMs. Our design is based on a principled integration of IC3 and LLMs. 
The symbolic component of \tool, referred to as the \emph{IC3 controller}, follows the IC3 procedure, 
decomposing invariant synthesis into focused blocking tasks that are concrete enough for an LLM to handle. The LLM, as the neural component, complements this process by providing protocol-level reasoning that IC3 alone lacks for TLA+ specifications, while the IC3 framework ensures that every proposed clause is validated and provides a completeness fallback.

More concretely, for each bad state exposed at the frontier frame, the IC3 controller performs a backward search through predecessor states in earlier frames. Once the search reaches a predecessor-free state, the LLM proposes protocol-level blocking clauses that exclude it. Each candidate clause is validated by the verification backend before being added to the current frame and then propagated forward. 
If no further bad states are found and the LLM fails to produce an admissible clause after repeated attempts, the solver falls back to an exact aggregate blocker that negates all accumulated bad states, preserving finite-instance completeness. The process terminates when two adjacent frames coincide, yielding a candidate inductive invariant. 
Notably, \tool does not impose restrictions on the input specification and requires no manually designed templates.

We implement the \tool design and evaluate it on 29 distributed protocols spanning consensus, leader election, lock services, transaction commit, firewall and client--server benchmarks. We compare against Endive~\cite{schultz2022kvy}, IC3PO~\cite{goel2021ic3po}, SWISS~\cite{hance2021swiss} and DistAI~\cite{yao2021distai}. Remarkably, \tool successfully discovers candidate invariants for all 29 protocols, including MongoLoglessDynamicRaft (MLDR), an industrial-scale Raft-based reconfiguration protocol on which none of the four compared tools report a solved result, as well as one complex Paxos consensus variant. In every solved case, we further prove, in TLAPS~\cite{chaudhuri2008tlaps}, that the discovered invariants are inductive for unbounded protocol instances, thereby establishing safety.

\medskip
\noindent\emph{Summary.} Our main contributions include:
\begin{itemize}[leftmargin=*]
    \item We propose a novel neuro-symbolic integration of IC3's frame-based decomposition with LLM-generated blocking clauses, turning globally intractable direct invariant synthesis into a sequence of focused per-state blocking tasks that are tractable for LLMs.
    \item We implement an IC3-based TLA+ tool, \tool, with an explicit IC3 controller and LLM-generated blocking clauses, requiring neither logical restriction on the input specification nor manual template design.
    \item We evaluate \tool on 29 distributed protocols and compare with Endive, IC3PO, SWISS, and DistAI. \tool discovers proven inductive invariants for all 29 benchmarks, including several that none of the compared tools solve.
\end{itemize}

\medskip
\noindent\emph{Roadmap.}
Section~\ref{sec:background} reviews background.
Section~\ref{sec:motivation} presents a motivating example.
Section~\ref{sec:appro} and Section~\ref{sec:impl} present the design and implementation details of \tool, respectively.
Section~\ref{sec:exper} reports evaluation results.
Section~\ref{sec:discuss} discusses case studies and limitations.
Section~\ref{sec:relate} surveys related work.
Section~\ref{sec:concl} concludes the paper and outlines future directions.

\section{Background}\label{sec:background}

This section introduces the three foundations that \tool builds on:
the TLA+ specification language (\S\ref{sec:bg-tlaplus}),
the IC3/PDR backward-blocking algorithm (\S\ref{sec:bg-ic3}),
and the TLC and Apalache model checkers (\S\ref{sec:bg-mc}).

\subsection{TLA+ Specification Language}\label{sec:bg-tlaplus}

TLA+ is a high-level specification language for modeling concurrent and distributed systems, based on the Temporal Logic of Actions~\cite{lamport2002specifying}. It models a system as a transition system over state variables, where each state assigns values to the variables, and each step is described by an action relating the current-state and next-state values.

In practice, a protocol specification typically introduces a tuple of variables \texttt{vars}, an initial-state predicate \texttt{Init}, and a next-state relation \texttt{Next} that captures the allowed transitions. The overall behavior is then written in the standard form \texttt{Init /\ [][Next]\_vars}, meaning that execution starts from a state satisfying \texttt{Init} and every step either follows \texttt{Next} or leaves the variables unchanged.

For distributed protocols, TLA+ specifications are usually written at the protocol level (rather than the implementation level), using sets, functions, and quantification over nodes, messages, or terms. Safety requirements are expressed as state predicates that must hold in every reachable state. A central verification task is therefore to find an inductive invariant that implies the target safety property, holds initially, and is preserved by \texttt{Next}. This is exactly the setting of our work: the protocol is given as a TLA+ transition system, and the challenge is to automatically strengthen the target safety property into an inductive invariant.

\subsection{IC3/PDR Intuition and Backward Blocking}\label{sec:bg-ic3}

IC3~\cite{bradley2011sat,een2011efficient} (also known as Property Directed Reachability, or PDR) is a frame-based, property-directed algorithm for proving safety properties of transition systems. We explain its key ideas informally because \tool builds directly on them.

The core idea is to incrementally construct an inductive invariant by maintaining a sequence of over-approximations of reachable states, called frames:
\[F_0\subseteq F_1\subseteq \cdots \]
where $F_0$ is the set of initial states, and each $F_i$ excludes known bad states.

The algorithm proceeds as follows. (1) It queries whether a bad state violating the safety property is reachable from the current frontier in one transition step. (2) If so, it traces the counterexample backward to find predecessor states. (3) It then blocks these states by adding clauses (lemmas) to the frames, preventing similar counterexamples. (4) These clauses are propagated forward to strengthen later frames.

The process continues until either a real counterexample to the property is found, or
two consecutive frames become identical ($F_k = F_{k+1}$), yielding an inductive invariant that proves safety.

\subsection{TLC and Apalache Model Checkers}\label{sec:bg-mc}

TLC~\cite{yu1999model} is the standard explicit-state model checker for TLA+. It explores the reachable state space of a TLA+ specification by exhaustively enumerating states and transitions, making it particularly effective for detecting safety violations in finite instances of distributed protocols. TLC is widely used in practice due to its simplicity and ability to produce concrete counterexample traces, which are invaluable for debugging protocol designs.

Apalache~\cite{konnov2019tla}, in contrast, is a symbolic model checker for TLA+ that leverages SMT solving. Instead of enumerating states explicitly, Apalache encodes the transition system and properties into logical constraints and uses an SMT solver to reason about sets of states symbolically. This approach enables more scalable analysis for certain classes of systems and supports bounded model checking, making it well-suited to the witness queries inside our IC3 loop and to exploring behaviors beyond the limits of explicit-state exploration.

TLC and Apalache represent two complementary verification paradigms for TLA+: explicit-state exploration for exhaustive bug finding on finite instances, and symbolic reasoning for more scalable, constraint-based analysis. They are both used in \tool, as described in \S\ref{sec:impl}.
 
\section{Motivating Example}\label{sec:motivation}

We use a two-phase commit protocol modeled in TLA+ to illustrate the challenge of inductive-invariant synthesis and the benefit of IC3 decomposition.

The protocol models $n$~nodes coordinating a commit-or-abort decision through seven set-valued variables (\greycode{vote\_yes}, \greycode{vote\_no}, \greycode{alive}, \greycode{go\_commit}, \greycode{go\_abort}, \greycode{decide\_commit}, \greycode{decide\_abort}) and a boolean \greycode{abort\_flag}.
Nodes vote yes or no; a node that votes no can immediately abort, since the outcome is already determined.
A coordinator broadcasts a commit signal (only when all nodes voted yes) or an abort signal (when some node voted no or failed); nodes then finalize their decisions accordingly.
Any node may fail at any time.
In this model, both voting no and failing set \greycode{abort\_flag}.
The safety property requires: (1)~no committed node coexists with an aborted node, (2)~every committed node implies all nodes voted yes, and (3)~every abort decision carries the abort flag.

\smallskip
\noindent\textbf{Why is safety not inductive?}
Safety alone is not an inductive invariant.
Consider the state $\greycode{go\_abort} = \{n_1\}$, $\greycode{abort\_flag} = \mathit{FALSE}$, with all other variables at their initial values.
This state satisfies Safety (no node has committed or aborted yet), but from it, node~$n_1$ can execute the \greycode{Abort} action (since $n_1 \in \greycode{go\_abort}$), producing $\greycode{decide\_abort} = \{n_1\}$ with $\greycode{abort\_flag} = \mathit{FALSE}$, violating the third safety conjunct.
An inductive invariant must include additional strengthening clauses that exclude such states.

\smallskip
\noindent\textbf{Direct LLM synthesis fails.}
We test a direct approach: the LLM (Claude Opus~4.6) generates a complete inductive invariant; a model checker verifies the induction condition; if it fails, the counterexample is fed back and the LLM tries again.
On this protocol, the LLM makes 44~attempts and times out without finding a correct invariant.
Each attempt produces a plausible candidate with 13--26 clauses, many individually correct (e.g., $\greycode{vote\_yes} \cap \greycode{vote\_no} = \{\}$ and $\greycode{decide\_commit} \subseteq \greycode{go\_commit}$).
However, the conjunction consistently fails the induction check because the LLM also includes overly strong clauses.
For instance, attempts frequently include $\greycode{go\_abort} \neq \{\} \Rightarrow \greycode{vote\_yes} \neq \mathit{Node}$, asserting that an abort signal is broadcast only when not all nodes voted yes.
This clause fails on reachable states of the simplified model, where crashed nodes do not recover: a node may vote yes and subsequently crash; the coordinator observes the failure and broadcasts an abort signal, producing a state where \greycode{go\_abort} is nonempty yet every node has voted yes.
The induction counterexample from the full candidate does not tell the LLM \emph{which} clause is wrong, so the LLM oscillates between variants without converging.

\smallskip
\noindent\textbf{IC3 decomposition.}
The failure suggests an alternative strategy: decompose the synthesis task so that each step targets a single concrete bad state. Interestingly, IC3's frame-based structure offers exactly this decomposition: clauses are organized into successive frames; each frontier query exposes one bad state, and the LLM only needs to propose clauses that block that state.

We apply this IC3-based approach to the same protocol.
The IC3 controller queries the frontier frame and finds a bad state: $\greycode{go\_abort} = \{n_1\}$, $\greycode{alive} = \{n_1\}$, with all other variables at their initial values.
Backward predecessor search confirms it is predecessor-free.
The LLM then produces 15~candidates targeting this state; the controller selects $\greycode{go\_abort} = \{\} \lor \greycode{go\_abort} = \mathit{Node}$, validates it, and admits it to the current frame.

The next frontier query reveals a new bad state, which is already covered by an existing clause in the pool ($\greycode{go\_commit} = \{\} \lor \greycode{go\_commit} = \mathit{Node}$), so an additional LLM call is not needed.
This process continues for 7~frontier rounds, of which only 2~require new LLM calls, with the remaining 5~reusing previously learned clauses.
After frontier blocking and clause pushing, during which all push failures are resolved by reusing previously learned clauses without additional LLM calls, adjacent frames converge, resulting in  a fixpoint of 11~strengthening clauses, automatically minimized to 8, obtained in 171~seconds with a total of 2~LLM calls.

Importantly, the IC3 controller, rather than the LLM, decides which clauses to keep. Candidate clauses are filtered through admission checks, and only validated clauses are propagated forward.
Incorrect or overly strong clauses are discarded during admission (rather than silently corrupting the invariant), so that IC3 compensates for LLM imprecision by construction.

Beyond mitigating LLM imprecision, the IC3 structure also provides a finite-instance completeness backstop. Namely, if no admissible LLM-guided clause can block the remaining bad states and no further bad states are found by Apalache, the solver falls back to an exact-checked aggregate blocker that excludes all existing bad states. In contrast, flat CTI-elimination approaches such as Endive~\cite{schultz2022kvy} do not provide this completeness guarantee.

\smallskip
\noindent\textbf{Takeaway.}
Direct synthesis requires the LLM to get all clauses correctly at once, and an induction failure over the full conjunction provides little guidance on which clause is failing. The IC3 decomposition circumvents the problem: each frontier query exposes a concrete bad state, the LLM proposes clauses to block that state, and the verification backend validates each clause individually before admission. This localisation of reasoning and feedback makes the search both more tractable and more robust, which underpins 
the design of~\tool.
 
\begin{figure*}[t]
  \centering
  \includegraphics[width=\linewidth,trim=9pt 98pt 28pt 101pt,clip]{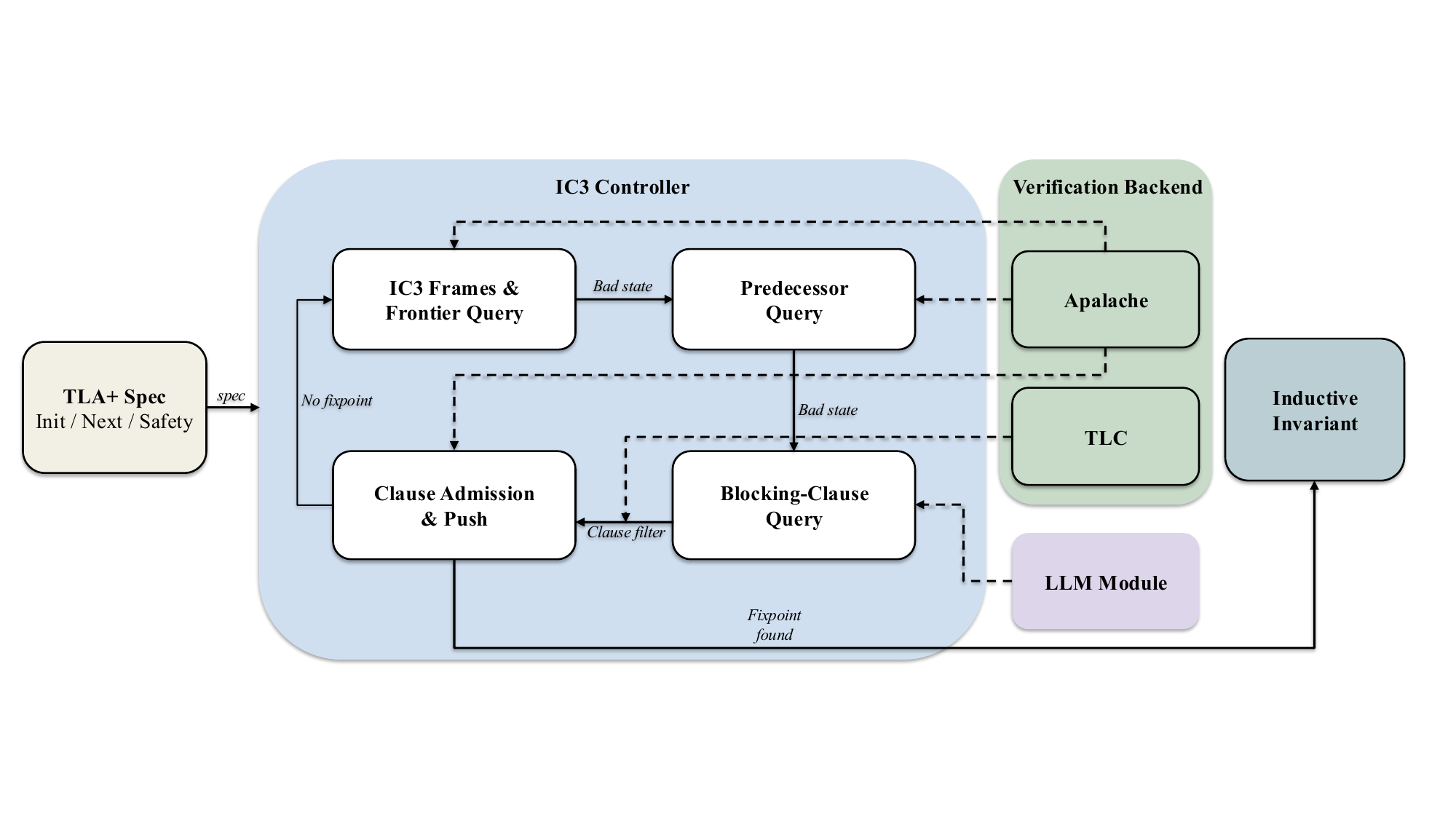}
  \caption{Overview of \tool. The IC3 loop iterates through four stages: frontier query, backward predecessor search, blocking-clause search, and clause push. They are coordinated by three components: (1) an IC3 controller that maintains frames and manages backward search from frontier bad states, (2) a verification backend where Apalache handles symbolic witness queries and TLC screens candidate blocking clauses against concrete reachable states, and (3) an LLM module that proposes candidate blocking clauses from protocol context and representative bad states.}
  \Description{Overview of IC3Syn. The figure shows a four-stage loop. The frontier query finds a bad state in the current deepest frame. Backward predecessor search traces it through shallower frames to a predecessor-free state. Blocking-clause search consults previously learned clauses and, if needed, queries the LLM for new candidates, which TLC screens against concrete reachable states before admission into IC3 frames. Clause push propagates admitted clauses toward deeper frames. The loop terminates when two adjacent frames coincide, yielding the candidate invariant.}
  \label{fig:overview}
\end{figure*}

\section{Design}\label{sec:appro}

Figure~\ref{fig:overview} presents an overview of \tool, which consists of three components: the \textbf{IC3 controller} that maintains frames and manages backward search from frontier bad states, the \textbf{verification backend} that handles property-directed witness queries and screens candidate blocking clauses against concrete reachable states, and the \textbf{LLM module} that proposes candidate blocking clauses from protocol context and representative failing states.

\subsection{IC3 Controller}

The IC3 controller maintains the explicit frame sequence $\mathit{Frames} = [\{\texttt{Init}\}, \{\texttt{Safety}\}]$, where $F_0 = \texttt{Init}$ and every frame $F_i$ with $i > 0$ begins from an internal \texttt{Safety} base that is then strengthened by admitted blocking clauses.
Each iteration proceeds through four stages.

\smallskip
\noindent\textbf{Frontier Query.}
The controller queries the current deepest frame for a \emph{frontier bad state}, i.e., a state that can lead to a safety violation in one transition.
If no such state exists, the frontier is already safe in one step and the controller proceeds to the push phase.
Otherwise, the bad state enters the backward predecessor search.

\smallskip
\noindent\textbf{Predecessor Query.}
The controller maintains a worklist of bad states, each paired with its frame level.
For each bad state, it issues a predecessor query one level shallower each time: if a predecessor exists, the bad state is replaced by its predecessor at the lower frame level and the search descends further; if no predecessor exists at a level above zero, the bad state is \emph{predecessor-free} and triggers blocking-clause search.
If the search reaches frame level~0, a real counterexample exists and the controller reports failure.

\smallskip
\noindent\textbf{Blocking-Clause Query.}
Once backward predecessor search reaches a predecessor-free bad state, the controller first consults previously learned clauses (i.e., a pool of all blocking clauses retained across the full IC3 process) for any that already block it.
Note that standard IC3 does not need such a store because clause derivation via SAT generalization is deterministic. However, with LLM-generated clauses, a clause originally learned for one bad state may later block a different bad state at a different frame level, so the pool avoids redundant LLM calls and enables cross-iteration reuse.

If existing clauses suffice, they are inserted into the relevant frames and the covered bad state is removed from the worklist without invoking the LLM.
Otherwise, the controller invokes the LLM module to propose candidate blocking clauses from the current batch of bad states.
Returned candidates are screened against concrete reachable states and retained for later reuse when they pass.
Clauses enter IC3 frames only after satisfying the initial-state implication and the target-frame admission condition, confirming that each clause is inductive relative to the frame's current contents.

If blocking fails, the bad state remains in the worklist and the controller re-queries the frontier, conjoining the negation of all known bad states, to obtain a fresh frontier bad state and try again.
If no new frontier bad states can be produced (meaning the known bad states already suffice to block the frontier), the controller negates all known bad states and adds the resulting negation to the frame as a last-resort blocking clause after exhausting LLM-guided clause search, preserving the finite-instance completeness backstop of IC3.
This guarantee distinguishes \tool from flat CTI-elimination approaches: even when the LLM fails to propose a useful clause, the IC3 loop still converges on the checked finite instance.

\smallskip
\noindent\textbf{Clause Push.}
Once the frontier is fully blocked, the controller appends a fresh frame and enters the push phase: each admitted clause is propagated toward deeper frames as far as the IC3 induction check allows.
A clause that cannot be pushed is simply non-inductive for the next frame; the solver makes one opportunistic blocking attempt on the push-failure counterexample, first consulting retained clauses and then issuing at most one additional LLM query if needed, but if no admissible clause is found, it moves on.
After propagation, the controller checks whether any two adjacent frames have become equal.
Equal adjacent frames constitute a fixpoint: the shared frame formula is extracted as the candidate invariant.

\subsubsection*{Key deviations from Standard IC3.}
Our design builds on IC3, which effectively transforms the global invariant synthesis problem into a sequence of localized, per-state blocking tasks. This structure aligns naturally with LLM-based clause generation, where each query can focus on a concrete sub-problem rather than the full invariant. However, \tool also features key deviations from standard IC3. 

The main adaptation lies in how blocking clauses are learned. In standard IC3, every predecessor-free bad state is blocked immediately by a SAT-derived Boolean cube generalization, a mechanical step that always succeeds.
In contrast, TLA+ specifications of distributed protocols involve sets, functions, and quantifiers that have no natural Boolean-cube representation, and a single bad state may not give the LLM sufficient context to infer the appropriate protocol-level condition.
To address this, \tool allows blocking to be deferred: when no admissible clause is found for a given bad state, the controller continues exploring the frontier, accumulating additional bad states in a shared worklist.
The resulting batch is then presented to the LLM, providing a richer context for generalization across multiple related failures rather than handling bad states in isolation. 

Moreover, \tool adds opportunistic blocking during the push phase.
When a clause fails to propagate to the next frame, the resulting push-failure counterexample exposes a bad state at the source frame.
Standard IC3 simply leaves the clause in its current frame, but \tool performs an additional blocking attempt: it first checks whether existing clauses suffice, and otherwise queries the LLM for a new clause targeting this push-failure state.
Any admissible clause is validated and incorporated under the same checks to guarantee soundness. If no such clause is found, the solver simply proceeds as usual, treating the push failure as a normal non-inductiveness event.

\subsection{Verification Backend and LLM Module} \label{sect:4.2}

\tool combines two model checkers as the verification backend, together with an LLM module, to support the IC3 controller.

\noindent\textbf{Symbolic witness queries (Apalache).}
The frontier and predecessor queries inside the IC3 loop are satisfiability problems: ``does there exist a state in the current frame that can step to a safety violation?'' and ``does there exist a predecessor of this bad state in the shallower frame?''
We hence use Apalache (cf.\ \S\ref{sec:bg-mc}),  whose SMT-based encoding handles quantified TLA+ state spaces without explicit enumeration. This is essential because the query spaces of frontier and predecessor witnesses can exceed the limits of explicit-state enumeration for some protocols.

\smallskip
\noindent\textbf{Concrete clause screening (TLC).}
Once the LLM proposes candidate blocking clauses, they must be validated against the protocol's reachable behavior.
We use TLC (cf.\ \S\ref{sec:bg-mc}) to screen candidates in batch on the configured finite instance by checking them against concrete reachable states.
This is a model-checking problem over a known, finite state space, where TLC's explicit enumeration is particularly effective.
Clauses which are falsified by reachable states are discarded early, and the corresponding counterexamples provide concrete diagnostic feedback to the LLM, steering subsequent proposals away from the same overgeneralization.
Moreover, TLC supports clause prioritization: for each valid clause, it counts how many current bad states are excluded and uses this metric to rank clauses for admission into the frame.

\smallskip
\noindent\textbf{LLM clause generation.}
The LLM module receives the full TLA+ specification along with the current batch of bad states, and proposes candidate blocking clauses expressed as quantified TLA+ predicates over protocol-level variables.
Unlike SAT-derived Boolean cubes that flip specific state assignments, these clauses capture protocol-level invariant conditions, such as ``every node in the quorum has voted'' or ``no committed log entry can be overwritten by a lower term'', allowing them to block entire families of bad states at once and accelerate convergence.

A key design choice is to query the LLM for individual blocking clauses scoped to specific bad states at a given frame level, rather than asking for a complete invariant.
This crucial decomposition creates focused, tractable sub-problem: the LLM needs only to propose predicates that exclude the given bad states while preserving reachable behavior, rather than discovering all clauses simultaneously.
Batching multiple bad states into a single call further strengthens generalization, enabling the LLM to detect
common structural patterns and to produce clauses that rule out families of violations rather than individual instances. 

The LLM also receives accumulated verification feedback (e.g., which prior clauses held on reachable states, which failed, and the specific falsifying states), steering it away from repeated overgeneralization and creating a productive verification--generation feedback loop.
 
\section{Implementation}\label{sec:impl}

We present the implementation details of \tool, including the IC3 controller algorithm, the selection of verification backends and their respective roles, and the design of LLM prompts and response handling for blocking clause generation.

\subsection{IC3 Main Loop}

{
  \setlength{\algomargin}{1em}
  \SetAlgoNoLine
  \LinesNumbered
  \SetAlFnt{\small}
  \begin{algorithm}[t]
    \caption{IC3 Loop with LLM-Generated Blocking Clauses\label{alg:main}}
    \KwIn{Initial condition \textup{\texttt{Init}}, safety property \textup{\texttt{Safety}}}
    \KwOut{Invariant $\mathit{Inv}$ or protocol is not safe \textsc{Unsafe}}
    $\mathit{F} \gets [\{\textup{\texttt{Init}}\}, \{\textup{\texttt{Safety}}\}]$\;
    $\mathit{clauseMemory} \gets \emptyset$\;
    \While{\textbf{true}}{
      \If{\textbf{not} $\textsc{BlockFrontier}(\mathit{F},\;\mathit{clauseMemory})$}{
        \KwRet{\textsc{Unsafe}}\;
      }
      $\textsc{PushClausesForward}(\mathit{F})$\;
      $j \gets \textsc{FindFixpoint}(\mathit{F})$\;
      \If{$j \neq \mathsf{nil}$}{
        $\mathit{Inv} \gets \textsc{ExtractInvariant}(F_j)$\;
        \KwRet{$\mathit{Inv}$}\;
      }
    }
  \end{algorithm}
}

Algorithm~\ref{alg:main} presents the complete IC3 control flow (corresponding
to the IC3 controller highlighted in blue in Figure~\ref{fig:overview}).
The controller initializes $\mathit{Frames} = [\{\texttt{Init}\}, \{\texttt{Safety}\}]$ and
maintains a global $\mathit{clauseMemory}$ that retains every filtered blocking clause
across the full IC3 process, allowing later iterations to reuse clauses
discovered during earlier frontier-blocking phases without re-querying the LLM.
Each iteration proceeds in three steps.

First, \textsc{BlockFrontier} performs the entire frontier-blocking phase for the
current deepest frame.
If backward search reaches frame level~0, it returns \textbf{false} and the controller reports \textsc{Unsafe}.

Second, once \textsc{BlockFrontier} has blocked all bad states reachable in one step from the
current frontier frame,
\textsc{PushClausesForward} appends a fresh frame and propagates each
admitted clause toward deeper frames by checking $F_k \wedge \texttt{Next} \Rightarrow c'$;
clauses that pass are pushed to the next frame.
An unpushable clause produces a push-failure pre-state, which triggers one
opportunistic blocking attempt against $\mathit{clauseMemory}$ and, if needed, at most one LLM query, subject to the standard
clause-admission check ($\mathit{Init} \Rightarrow c$ and $F_{k-1} \wedge c \wedge \texttt{Next} \Rightarrow c'$); if no admissible
clause is found, the solver simply moves on and the pre-state is discarded.
Any new clause added to a frame triggers re-pushing from that frame.

Third, \textsc{FindFixpoint} scans adjacent frame pairs and returns the index~$j$
of the first pair satisfying $F_j = F_{j+1}$; if found, the controller extracts
and returns the candidate invariant.

\subsection{Frontier Frame Bad-State Blocking}

{
  \setlength{\algomargin}{1em}
  \SetAlgoNoLine
  \LinesNumbered
  \SetAlFnt{\small}
  \begin{algorithm}[t]
    \caption{\textsc{BlockFrontier}: Frontier Bad-State Blocking\label{alg:blockfrontier}}
    \KwIn{Frame sequence $\mathit{F}$, clause memory $\mathit{clauseMemory}$}
    \KwOut{\textbf{true} if no more bad states}
    $\mathit{badState} \gets \textsc{FrontierQuery}(F)$\; \label{line1}
    $\mathit{badStates} \gets \{\mathit{badState}\}$\;
    \While{$\mathit{badStates} \neq \emptyset$}{
      $\mathit{badState} \gets \textsc{PredecessorQuery}(\mathit{badState},\;\mathit{F})$\; \label{line4}
      $\mathit{badStates} \gets \mathit{badStates} \cup \{\mathit{badState}\}$\;
      \If{$\mathit{frameLevel} = 0$}{
        \KwRet{\textbf{false}}
      }
      $\mathit{clauses} \gets \textsc{AdmitClauses}(\mathit{clauseMemory},\;\mathit{badStates})$\;
      \If{$\mathit{clauses} \neq \emptyset$}{
        \textbf{continue}\;
      }
      $\mathit{candidateClauses} \gets \textsc{LearnClauses}(\mathit{badStates},\;\mathit{F})$\;
      $\mathit{candidateClauses} \gets \textsc{Filter}(\mathit{candidateClauses},\;\mathit{F})$\;
      $\mathit{clauseMemory} \gets \mathit{clauseMemory} \cup \mathit{candidateClauses}$\;
      $\mathit{clauses} \gets \textsc{AdmitClauses}(\mathit{clauseMemory},\;\mathit{badStates})$\;
      $\mathit{badStates} \gets \textsc{RemoveBlocked}(\mathit{badStates},\;\mathit{clauses})$\;
      $\mathit{badState}  \gets \textsc{FrontierQuery}(\mathit{F},\;\mathit{badStates})$\; \label{line18}
      $\mathit{badStates} \gets \mathit{badStates} \cup \{\mathit{badState}\}$\;
    }
    \KwRet{\textbf{true}}\;
  \end{algorithm}
}

Algorithm~\ref{alg:blockfrontier} details the frontier-blocking phase.
It queries the current deepest frame for a state from which one transition reaches a Safety-violating successor and adds this pre-state to the
bad-state worklist~$\mathit{badStates}$.
A predecessor query then checks whether $F_{k-1} \wedge \texttt{Next}
  \Rightarrow \neg\mathit{badState}'$; if a predecessor is found, the bad state is
replaced by its predecessor at the lower frame level, and the search descends
further.
If the search reaches frame level~0, a real counterexample path to a safety violation exists and the procedure returns~\textbf{false}.
If no predecessor exists at any frame level above zero, the bad state
is \emph{predecessor-free} and triggers blocking-clause search.
Predecessor checks are performed per bad state.

Before invoking the LLM, the loop first calls \textsc{AdmitClauses} on the
current clause memory; if an existing clause can cover the current bad
states and is admissible, the loop continues without querying the LLM.
Only when no retained clause is admissible does the loop call
\textsc{LearnClauses}, which submits the
protocol specification, the current batch of bad states, verification histories of prior LLM-generated clauses, and
existing clauses in the current frame to the LLM.
If this still does not yield an admissible clause, the bad state remains in
the worklist, and the next iteration attempts to discover additional bad states
and generate new clauses.

New candidate clauses are filtered via \textsc{Filter} and merged into
$\mathit{clauseMemory}$: TLC checks each candidate against the accumulated
reachable states and discards those that are falsified, since they cannot
appear in the final invariant and would only waste subsequent verification attempts (cf.\ \S\ref{sect:ver}).

Moreover, TLC checks clauses in $\mathit{clauseMemory}$ against the current bad states and retains only those that block at least one bad state;
among the retained clauses, those covering more bad states are preferred, and when coverage is equal, shorter clauses are preferred.
\textsc{AdmitClauses} is then called again on the updated memory.
The admission check verifies $\mathit{Init} \Rightarrow c$ and
$F_{\mathit{k}-1} \wedge c \wedge \texttt{Next} \Rightarrow c'$ for each
candidate~$c$, inserting admitted clauses into frames from level 1 up to the current level. Once any clause is successfully admitted, the current blocking attempt for this bad state terminates.
After admission, \textsc{RemoveBlocked} removes from $\mathit{badStates}$ every
bad state covered by the newly admitted clauses;
\textsc{FrontierQuery} then re-queries the frontier,
excluding already-seen pre-states, so that newly exposed frontier bad states
re-enter the same round.

\subsection{Verification Backend} \label{sect:ver}

As stated in \S~\ref{sect:4.2}, Apalache and TLC serve complementary roles in \tool. 

Apalache handles all symbolic queries in the IC3 loop: the frontier query (lines \ref{line1} and \ref{line18} in Algorithm~\ref{alg:blockfrontier}), the predecessor query (line \ref{line4}), clause admission ($\mathit{Init} \Rightarrow c$ and $F_{\mathit{k}-1} \wedge c \wedge \texttt{Next} \Rightarrow c'$), and clause push ($F_k \wedge \texttt{Next} \Rightarrow c'$).
All rely on its symbolic one-step reachability encoding, which avoids state enumeration and returns concrete witness states directly.

TLC screens candidate clauses in batches: the entire candidate pool is validated against reachable states in a single model-checking pass rather than checking each clause individually.
For each surviving clause, TLC counts how many current bad states it excludes; this coverage count ranks candidates before admission.

Apalache is chosen for witness queries because the search spaces of frontier and predecessor queries can exceed the limits of explicit-state enumeration for some protocols, whereas TLC's batched explicit-state checking is more efficient for screening concrete clause candidates.

\subsection{LLM Clause Generator}

Each prompt includes (1) the TLA+ protocol specification, (2) a batch of current bad
states, (3) quantifier templates extracted from the specification, (4) the persistent cross-round clause memory (whose clauses should not be repeated),
and (5) per-clause
observations tracking which candidates have previously held or failed on
reachable states.
The prompt requests complete TLA+ clauses as a JSON batch response. 
(Prompt~\hyperref[fig:prompt]{1} exemplifies a concrete template with minor simplifications for the sake of presentation.)

The LLM response is parsed as a JSON array, from which individual clauses are extracted,
normalized, and syntax-checked with SANY~\cite{lamport2002specifying}. The clauses are then filtered with TLC against reachable states.
Together, SANY syntax checking and TLC-based screening serve as additional filters, eliminating malformed or already-falsified candidates early.

Note that common scoping mistakes, such as reusing a model constant as a quantifier-bound
variable, are automatically detected and repaired  during normalization.
Quantifier templates are extracted directly from the specification and deduplicated modulo
variable renaming, providing the LLM with protocol-specific quantifier
structure without requiring manual template design.

\begin{figure}[t]
  \begin{definedbox}{Blocking-Clause Generation Prompt}
    \raggedright\small
    Generate TLA+ inductive invariant clauses that exclude the counterexample states while capturing structural relationships in the specification.

    \begin{enumerate}[leftmargin=1.5em,itemsep=1pt,topsep=2pt]
      \item Complete, self-contained TLA+ expressions; no primed variables or temporal operators.
      \item Use nested quantifiers for dependent domains, reusing the provided quantifier templates.
      \item Generalize from counterexamples; prefer structural clauses over witness-specific exclusions.
      \item Use only identifiers and operators present in the specification or counterexample states.
      \item Do not repeat existing clauses or return any clause listed under known false.
      \item Use known true clauses as structural guidance only; try strengthening them into tighter variants.
      \item For each known false entry, try loosening the clause by folding the violating state into its quantified or value space.
    \end{enumerate}

    \textbf{Specification:} \texttt{\{spec\_content\}}\\
    \textbf{Counterexample states:} \texttt{\{bad\_states\}}\\
    \textbf{Quantifier templates:} \texttt{\{quantifier\_templates\}}\\
    \textbf{Existing clauses:} \texttt{\{existing\_clauses\}}\\
    \textbf{Known true on finite reachable states:} \texttt{\{known\_true\_clauses\}}\\
    \textbf{Known false on finite reachable states:} \texttt{\{known\_false\_clauses\}}\\[2pt]
    \textbf{Output:}
    JSON array of 10--15 clauses:
    \texttt{[\{"clause\_name": "...", "clause": "TLA+ expr"\}]}
  \end{definedbox}
  \Description{Prompt template for LLM blocking-clause generation.}
  \label{fig:prompt}
\end{figure}
 
\section{Evaluation}\label{sec:exper}

We present an empirical evaluation of \tool with emphasis on the following research questions (\textbf{RQs}).

\begin{description}
    \item[\textbf{RQ1.}] How effective is \tool in inductive-invariant synthesis, and how does it compare with state-of-the-art tools?

    \item[\textbf{RQ2.}] What is the contribution of the IC3 architecture to \tool's effectiveness?
\end{description}

\subsection{Benchmark and Setup}

We evaluate \tool on 29 distributed protocols spanning consensus, leader election, lock services, transaction commit variants~\cite{gray2006consensus}, firewalls, client--server protocols, Paxos variants~\cite{lamport1998parttime,lamport2001paxos}, and a larger production-style case study (MongoLoglessDynamicRaft). The suite draws primarily from the Endive benchmark collection and extends it with an additional hard Paxos variant sourced from the IC3PO repository and adapted to TLA+.
Endive is directly comparable on the 28 shared TLA+ protocols; IC3PO~\cite{goel2021ic3po}, SWISS~\cite{hance2021swiss} and DistAI~\cite{yao2021distai} serve as additional baselines wherever a matching benchmark is available in their respective input languages.

We impose a 600-second timeout per tool on all protocols; for \texttt{pyv-paxos-epr} and \texttt{mldr}, we extend the limit to 2~hours, as Apalache queries on these two protocols are substantially more expensive. The 2-hour extended limit also applies to all tools.

All reported time is end-to-end wall-clock runtime per protocol, measured on the same machine.
All experiments were conducted on a MacBook Pro with an Apple M4 Pro chip (12 cores: 8 performance and 4 efficiency) and 48\,GB of memory.

\subsection{RQ1: Effectiveness and Comparison}

Table~\ref{tab:tool-comparison} provides a full per-protocol breakdown.
We compare \tool against four prior state-of-the-art automated invariant inference tools: Endive~\cite{schultz2022kvy}, IC3PO~\cite{goel2021ic3po}, SWISS~\cite{hance2021swiss}, and DistAI~\cite{yao2021distai}. Because Endive requires expert-curated quantifier and predicate templates, we do not report Endive results on the new benchmark protocol (shown as n/a in Table~\ref{tab:tool-comparison}).

\smallskip
\noindent\textbf{Baseline tools.}
Endive~\cite{schultz2022kvy} is the first invariant inference tool to operate directly on TLA+ specifications, using a flat CTI-elimination loop with a manually curated predicate pool; this design requires protocol-specific insight to craft the quantifier and predicate pool, making it hard for non-experts to apply.
IC3PO~\cite{goel2021ic3po} extends IC3/PDR with symmetric incremental induction over indistinguishable domain sorts, operating on Ivy and mypyvy specifications in decidable first-order fragments; it is highly effective when protocol correctness aligns with strong structural regularity but less general for invariants involving ordering, arithmetic, or non-symmetric relational structure.
SWISS~\cite{hance2021swiss} performs symmetry-aware exhaustive search over template-defined candidate invariant spaces with counterexample-based pruning.
DistAI~\cite{yao2021distai} takes a data-driven approach, enumerating candidate invariants from simulation traces and monotonically weakening them until an SMT solver confirms inductiveness; it is restricted to universally quantified invariant formulas.

\begin{table*}[!tp]
    \centering
    \begin{threeparttable}
        \caption{Comparison with the default \tool configuration (Claude Opus~4.6), a direct-LLM ablation (LLM-alone), a weaker-LLM ablation with Gemini~3 Flash (LLM-var), and prior state-of-the-art automated invariant inference tools. ``Inv'' reports the number of clauses in the discovered invariant (excluding \texttt{TypeOK}, including the safety-property clauses), which partially reflects the complexity of subsequent proof writing. ``TO'' = timeout, ``err'' = tool error, ``fail'' = terminated without finding an invariant, ``n/a'' = no comparable result.}
        \label{tab:tool-comparison}
        \small
        \setlength{\tabcolsep}{3.5pt}
        \begin{tabular}{@{}rl|rr|rr|rr|rr|rr||rr|rr@{}}
            \toprule
               &                                & \multicolumn{2}{c|}{\textbf{\tool}} & \multicolumn{2}{c|}{\textbf{Endive}} & \multicolumn{2}{c|}{\textbf{IC3PO}} & \multicolumn{2}{c|}{\textbf{SWISS}} & \multicolumn{2}{c||}{\textbf{DistAI}} & \multicolumn{2}{c|}{\textbf{LLM-alone}} & \multicolumn{2}{c@{}}{\textbf{LLM-var}}                                                                                                                                              \\
            \cmidrule(lr){3-4}\cmidrule(lr){5-6}\cmidrule(lr){7-8}\cmidrule(lr){9-10}\cmidrule(lr){11-12}\cmidrule(lr){13-14}\cmidrule(lr){15-16}
            \# & Protocol                       & Time                                & \phantom{T}Inv                       & Time                                & \phantom{T}Inv                      & Time                                  & \phantom{T}Inv                          & Time                                    & \phantom{T}Inv            & Time                       & \phantom{T}Inv & Time                    & \phantom{T}Inv & Time & \phantom{T}Inv \\
            \midrule
            \midrule
            1  & tla-consensus                  & 1                                   & 2                                    & 0                                   & 2                                   & 0                                     & 1                                       & 0                                       & 3                         & 1                          & 0              & 13                      & 3              & 2    & 2              \\
            2  & tla-tcommit                    & 1                                   & 1                                    & 1                                   & 1                                   & 0                                     & 2                                       & fail                                    & 1                         & 1                          & 6              & 12                      & 4              & 2    & 1              \\
            3  & i4-lock-server                 & 60                                  & 2                                    & 3                                   & 2                                   & 0                                     & 2                                       & 0                                       & 2                         & 1                          & 1              & 9                       & 3              & 32   & 2              \\
            4  & ex-quorum-leader-election      & 41                                  & 2                                    & 6                                   & 2                                   & 1                                     & 4                                       & 32                                      & 6                         & 1                          & 7              & 13                      & 7              & 56   & 2              \\
            5  & pyv-toy-consensus-forall       & 42                                  & 3                                    & 8                                   & 3                                   & 2                                     & 4                                       & 13                                      & 5                         & 2                          & 6              & 10                      & 6              & 68   & 3              \\
            6  & tla-simple                     & 50                                  & 2                                    & 4                                   & 2                                   & 2                                     & 4                                       & 25                                      & 8                         & \multicolumn{2}{c||}{err}  & 15             & 7                       & 52             & 3                     \\
            7  & ex-lockserv-automaton          & 92                                  & 9                                    & 9                                   & 9                                   & 2                                     & 9                                       & fail                                    & 1                         & 1                          & 12             & 12                      & 12             & 222  & 9              \\
            8  & tla-simpleregular              & 45                                  & 3                                    & 4                                   & 4                                   & 1                                     & 4                                       & 31                                      & 21                        & \multicolumn{2}{c||}{err}  & 331            & 7                       & 59             & 2                     \\
            9  & pyv-sharded-kv                 & 117                                 & 6                                    & 132                                 & 6                                   & 1                                     & 6                                       & \multicolumn{2}{c|}{TO}                 & 1                         & 15                         & 56             & 10                      & 110            & 5                     \\
            10 & pyv-lockserv                   & 90                                  & 9                                    & 14                                  & 9                                   & 2                                     & 9                                       & \multicolumn{2}{c|}{TO}                 & 1                         & 12                         & 106            & 10                      & 303            & 8                     \\
            11 & tla-twophase                   & 165                                 & 9                                    & 19                                  & 10                                  & 4                                     & 11                                      & fail                                    & 1                         & 12                         & 305            & \multicolumn{2}{c|}{TO} & 318            & 9                     \\
            12 & i4-learning-switch             & 434                                 & 12                                   & \multicolumn{2}{c|}{TO}             & 11                                  & 9                                     & 551                                     & 13                                      & 13                        & 31                         & 16             & 14                      & 462            & 10                    \\
            13 & ex-simple-decentralized-lock   & 35                                  & 3                                    & 17                                  & 4                                   & 3                                     & 4                                       & 7                                       & 3                         & 7                          & 16             & \multicolumn{2}{c|}{TO}              & 43   & 3              \\
            14 & i4-two-phase-commit            & 171                                 & 11                                   & 30                                  & 13                                  & 2                                     & 11                                      & 6                                       & 14                        & 6                          & 66             & \multicolumn{2}{c|}{TO} & 209            & 11                    \\
            15 & pyv-consensus-wo-decide        & 92                                  & 5                                    & 52                                  & 8                                   & 75                                    & 6                                       & 38                                      & 8                         & \multicolumn{2}{c||}{err}  & 15             & 9                       & 243            & 7                     \\
            16 & pyv-consensus-forall           & 383                                 & 6                                    & 96                                  & 8                                   & 377                                   & 6                                       & 36                                      & 11                        & \multicolumn{2}{c||}{err}  & 11             & 9                       & 198            & 6                     \\
            17 & pyv-learning-switch            & 315                                 & 6                                    & \multicolumn{2}{c|}{TO}             & \multicolumn{2}{c|}{TO}             & \multicolumn{2}{c|}{TO}               & \multicolumn{2}{c||}{err}               & \multicolumn{2}{c|}{TO}                 & 227                       & 7                                                                                                              \\
            18 & pyv-sharded-kv-no-lost-keys    & 188                                 & 6                                    & 5                                   & 2                                   & 0                                     & 2                                       & 2                                       & 4                         & \multicolumn{2}{c||}{fail} & 10             & 3                       & 101            & 6                     \\
            19 & ex-naive-consensus             & 55                                  & 4                                    & 15                                  & 4                                   & 2                                     & 4                                       & \multicolumn{2}{c|}{TO}                 & 1                         & 1                          & 16             & 7                       & 129            & 4                     \\
            20 & pyv-client-server-ae           & 153                                 & 2                                    & 19                                  & 2                                   & 0                                     & 2                                       & 17                                      & 5                         & \multicolumn{2}{c||}{err}  & 9              & 3                       & 36             & 2                     \\
            21 & ex-simple-election             & 46                                  & 3                                    & 10                                  & 4                                   & 1                                     & 4                                       & \multicolumn{2}{c|}{TO}                 & \multicolumn{2}{c||}{err} & 10                         & 5              & 147                     & 4                                      \\
            22 & pyv-toy-consensus-epr          & 36                                  & 3                                    & 10                                  & 4                                   & 2                                     & 4                                       & 10                                      & 5                         & \multicolumn{2}{c||}{err}  & 9              & 6                       & 80             & 4                     \\
            23 & ex-toy-consensus               & 34                                  & 2                                    & 4                                   & 2                                   & 3                                     & 3                                       & \multicolumn{2}{c|}{TO}                 & \multicolumn{2}{c||}{err} & 11                         & 4              & 44                      & 2                                      \\
            24 & pyv-client-server-db-ae        & 90                                  & 4                                    & \multicolumn{2}{c|}{TO}             & 23                                  & 7                                     & 60                                      & 15                                      & \multicolumn{2}{c||}{err} & 19                         & 9              & 178                     & 5                                      \\
            25 & pyv-firewall                   & 35                                  & 2                                    & 24                                  & 5                                   & 2                                     & 3                                       & fail                                    & 1                         & \multicolumn{2}{c||}{err}  & 12             & 3                       & 68             & 2                     \\
            26 & ex-majorityset-leader-election & 40                                  & 5                                    & 27                                  & 5                                   & 14                                    & 6                                       & 24                                      & 11                        & \multicolumn{2}{c||}{err}  & 23             & 9                       & 62             & 5                     \\
            27 & pyv-consensus-epr              & 122                                 & 6                                    & 207                                 & 8                                   & 68                                    & 6                                       & 26                                      & 8                         & \multicolumn{2}{c||}{err}  & 11             & 9                       & 134            & 6                     \\
            28 & mldr                           & 4525                                & 16                                   & \multicolumn{2}{c|}{TO}             & \multicolumn{2}{c|}{TO}             & \multicolumn{2}{c|}{TO}               & \multicolumn{2}{c||}{err}               & \multicolumn{2}{c|}{TO}                 & \multicolumn{2}{c@{}}{TO}                                                                                                                  \\
            29 & pyv-paxos-epr                  & 1806                                & 9                                    & \multicolumn{2}{c|}{n/a}            & \multicolumn{2}{c|}{TO}             & \multicolumn{2}{c|}{TO}               & \multicolumn{2}{c||}{err}               & \multicolumn{2}{c|}{TO}                 & \multicolumn{2}{c@{}}{TO}                                                                                                                  \\
            \bottomrule
        \end{tabular}
    \end{threeparttable}
\end{table*}

\smallskip
\noindent\textbf{Effectiveness.}
Across the 29 protocols in Table~\ref{tab:tool-comparison}, \tool solves all 29.
The hardest case is \texttt{mldr} (MongoLoglessDynamicRaft), which takes 4525 seconds.
Much of the time is spent on the verification backend, where a single inductiveness check can exceed 400 seconds.
For all 29 solved benchmarks, TLAPS confirms that the discovered invariants are inductive for unbounded protocol instances, thereby establishing the corresponding safety arguments. This proof-writing task is mainly done by agents with minimal human intervention.

\smallskip
\noindent\textbf{Comparison with Endive.}
\tool eliminates Endive's reliance on manual predicate-pool design by replacing predicate enumeration with LLM-based clause generation guided by concrete counterexample states. This design makes \tool fully automated on new specifications without requiring protocol-specific templates.

On the 28 shared TLA+ protocols, \tool decisively outperforms Endive, solving 28 while Endive solves 24. Beyond the automation gap, Endive's flat CTI-elimination loop lacks an IC3-style completeness backstop. By retaining the IC3 controller, \tool preserves the finite-instance completeness guarantee while shifting clause discovery to LLM-generated blocking clauses.

\smallskip
\noindent\textbf{Comparison with IC3PO, SWISS and DistAI.}
IC3PO solves 26 of the 29 benchmarks and is often faster than \tool on the protocols it covers, but it times out on the three hardest cases (\texttt{pyv-learning-switch}, \texttt{mldr}, and \texttt{pyv-paxos-epr}). As discussed above, its symmetry-based method is less general for protocols requiring richer ordering or arithmetic reasoning.
SWISS solves 17 of the 29 comparable benchmarks, timing out on 8 and terminating without an invariant on 4.
DistAI is the weakest baseline: several rows end in errors, and \texttt{pyv-sharded-kv-no-lost-keys} terminates without a valid invariant despite returning a large conjunct set.

\subsection{RQ2: Contribution of the IC3 Architecture}

To isolate the contribution of the IC3 loop, we evaluate two ablation variants: (1)~a direct-LLM baseline that uses the same underlying model (Claude Opus~4.6) but omits the IC3 frame structure entirely, and (2)~a weaker-LLM variant that replaces Claude Opus~4.6 with Gemini~3 Flash inside the full \tool IC3 controller.

\smallskip
\noindent\textbf{Direct-LLM baseline.}
Given the TLA+ specification and the safety property, the baseline prompts the LLM to propose a complete inductive invariant, i.e., a single conjunctive formula that must explicitly include \texttt{TypeOK} and the safety property. The candidate invariant is then checked for syntactic correctness as well as proof obligations, i.e., \texttt{Init} $\Rightarrow$ \texttt{Inv} and \texttt{Inv} $\land$ \texttt{Next} $\Rightarrow$ \texttt{Inv'}.
If the candidate fails due to a syntax error, violation of the initialization condition, or lack of inductiveness, the failure is reported back to the LLM, and a new candidate is requested. This process iterates until a valid candidate is found or a per-benchmark time budget is exhausted (600\,s by default; 2\,h for \texttt{mldr} and \texttt{pyv-paxos-epr}).

On the same 29 benchmarks, the direct-LLM baseline solves 23 instances and times out on 6.
The six timed-out benchmarks are \texttt{ex-simple-decentralized-lock}, \texttt{tla-twophase}, \texttt{i4-two-phase-commit}, \texttt{pyv-learning-switch}, \texttt{mldr}, and \texttt{pyv-paxos-epr}, which are among the most challenging cases in the full \tool suite. These results indicate that an LLM alone can solve many of the simpler benchmarks, but the harder instances require the IC3 structure to effectively decompose the invariant inference task.

\smallskip
\noindent\textbf{Weaker LLM with IC3.}
The last column of Table~\ref{tab:tool-comparison} reports the result where the default Claude Opus~4.6 is replaced with Gemini~3 Flash. Under this substitution, \tool solves 27 of the 29 benchmarks, timing out on \texttt{mldr} and \texttt{pyv-paxos-epr}.

Notably, the direct-LLM baseline using Claude Opus~4.6 alone fails to solve \texttt{tla-twophase}, whereas the weaker Gemini~3 Flash paired with \tool's IC3 controller succeeds on the same protocol. This comparison highlights that the IC3 decomposition can compensate for limitations of the underlying model: the frame structure, clause admission, and backward predecessor search provide enough guidance for even a weaker model to produce useful blocking clauses, when each query is focused on a single concrete bad state.

\medskip
\noindent\textbf{Summary.}
The ablation studies demonstrate that the IC3 architecture is the key contributor to \tool's effectiveness. The frame-based decomposition and clause admission checks mitigate variability in clause quality, allowing a weaker model to solve a substantial fraction of benchmarks, while the direct-LLM baseline---even with the stronger model---fails on the harder protocols that require structured, incremental invariant construction.
 
\section{Discussion}\label{sec:discuss}

\subsection{Case Study: MongoLoglessDynamicRaft (MLDR)}

MongoLoglessDynamicRaft (MLDR) is a production-grade Raft variant with dynamic reconfiguration, modeling the core consensus mechanism used in MongoDB's replication layer. The protocol maintains five state variables per server---\greycode{currentTerm}, \greycode{state}, \greycode{configVersion}, \greycode{configTerm}, and \greycode{config}---and supports four actions: leader election via quorum voting, configuration propagation, dynamic reconfiguration, and term updates. The safety property requires at most one primary per term (\greycode{OnePrimaryPerTerm}).

\smallskip
\noindent\textbf{Invariant discovery.}
\tool discovers a 15-conjunct strengthening invariant (16 clauses in total, including safety) in 4,525 seconds, making MLDR the most difficult benchmark in our evaluation. The discovered conjuncts fall into three categories:

\begin{enumerate}[leftmargin=*]
       \item \textbf{Single-server constraints} (8 clauses): Well-formedness conditions scoped to individual servers, such as $\forall s \in \mathit{Server} : \mathit{configVersion}[s] \geq 1$ and $\forall s \in \mathit{Server} : \mathit{state}[s] = \mathit{Primary} \Rightarrow \mathit{configTerm}[s] = \mathit{currentTerm}[s]$. These are discovered quickly in early IC3 rounds.

       \item \textbf{Cross-server relationships} (4 clauses): Properties linking pairs of servers, including configuration uniqueness:
             \begin{align*}
                     & \forall s, t \in \mathit{Server} :                               \\
                     & \quad \mathit{configVersion}[s] = \mathit{configVersion}[t]      \\
                     & \quad {}\wedge\; \mathit{configTerm}[s] = \mathit{configTerm}[t] \\
                     & \quad \Rightarrow \mathit{config}[s] = \mathit{config}[t]
             \end{align*}
             and quorum overlap:
             \begin{align*}
                     & \forall s, t \in \mathit{Server} :                                        \\
                     & \quad \mathit{QuorumsOverlap}(\mathit{config}[s], \mathit{config}[t])     \\
                     & \quad {}\lor\; \mathit{ConfigDisabled}(s) \lor \mathit{ConfigDisabled}(t)
             \end{align*}

       \item \textbf{Nested-quantifier quorum properties} (3 clauses): The hardest clauses, requiring alternating $\forall\exists$ quantifiers over servers and quorums with cross-variable comparisons. One such clause is:
             \begin{align*}
                     & \forall s, t \in \mathit{Server} :                                           \\
                     & \quad \forall Q \in \mathit{QuorumsAt}(s) :                                  \\
                     & \quad \exists n \in Q : \mathit{currentTerm}[n] \geq \mathit{currentTerm}[t] \\
                     & \quad \lor\; \mathit{IsNewerConfig}(n, s)
             \end{align*}
             In plain language: for any two servers $s$ and $t$, every quorum in $s$'s configuration has a witness node $n$ that is either at least as current as $t$ in term number, or running a newer configuration than $s$. This prevents any quorum from becoming uniformly stale with respect to any other server's term---a necessary condition for reconfiguration safety. A second clause specializes this to competing primaries: if two primaries coexist with differing terms, every quorum of the lower-term primary must contain a node whose term already exceeds that primary's, preventing the stale primary from reconstituting its original election quorum. Together with a simpler quorum-level clause for single primaries, these three clauses form the crux of the reconfiguration safety argument.
\end{enumerate}

\smallskip
\noindent\textbf{Why is MLDR hard?}
The difficulty is not in obtaining counterexamples, but in eliminating residual bad states after the structural clauses have been added. The IC3 loop quickly removes most early-round counterexamples with local constraints, then faces a small residual set that can only be closed by a strong global property rather than a local state fact.
The nested-quantifier clauses require the LLM to infer cross-variable relationships (e.g., $\mathit{currentTerm}[n] \geq \mathit{currentTerm}[t] \lor \mathit{IsNewerConfig}(n, s)$) from individual counterexample states that typically expose only a few concrete node configurations. The quorum-level existential quantification is especially challenging because the blocking clause needs a witness \emph{set} with a specific structural property, not merely a simple universal constraint. Much of the  runtime (4,525 seconds) is spent on the verification backend, where a single inductiveness query can exceed 400 seconds.

\smallskip
\noindent\textbf{Comparison with baselines.}
Table~\ref{tab:tool-comparison} shows that Endive, IC3PO, SWISS and DistAI all fail on MLDR. Our local Endive run does not solve this protocol, even with a manually engineered predicate pool of 38 predicates covering configuration terms, versions, quorum overlap, and the protocol's key operators. Reproducing the archived Endive result appears to require an expert-designed quantifier and predicate pool that already encodes deep protocol knowledge, precisely the kind of manual effort that \tool eliminates.

\smallskip
\noindent\textbf{TLAPS proof.}
For formal verification beyond the checked finite instance, we prove the discovered invariant inductive for unbounded protocol instances using TLAPS. We first simplify the 15-conjunct output by iterating from older clauses to newer ones: at each step, we remove one clause and use Apalache to re-check inductiveness on the same finite instance. Note that it is fully automatic once the inductive invariant is available. This eliminates 8 redundant clauses, yielding a minimal 7-conjunct strengthening clauses. Such redundancy is expected: during the IC3 loop, clauses are introduced to strengthen intermediate frames and block specific bad states, but once stronger clauses are admitted, earlier weaker ones can be removed in a final simplification pass.

Recent work on agent-based system verification has found that automatically discovering inductive invariants remains beyond the reach of current LLMs~\cite{yang2025verusage}. However, once \tool provides the discovered invariants, the subsequent TLAPS proofs become tractable for LLM agents. In our evaluation, a proof of over 3,000 lines (MLDR) was primarily completed by LLM agents with minimal human intervention, and not a single line of proof was directly written by a human.

Concretely, an LLM-based agent first constructs a proof scaffold according to a set of structural rules, such as dividing proofs into per-clause and per-action obligations. A dedicated proof-advancement skill was then deployed on top of Codex's automation layer, which continuously attempts to discharge remaining proof obligations until completion. Integrating this agent-driven TLAPS workflow into a fully automated end-to-end pipeline is ongoing work (cf.\ \S~\ref{sec:concl}).

\subsection{Limitations}
We briefly discuss the limitations of our work and how we mitigate them.

\noindent\textbf{Finite-instance validation.}
The \tool loop validates inductiveness of invariants only on finite instances. Both TLC and Apalache reason about bounded state spaces, so the IC3 fixpoint on the checked instance does not automatically generalize to arbitrary parameter configurations.

In the current work, we close this gap using TLAPS: for all 29 solved benchmarks, TLAPS proves that the discovered invariants are inductive for unbounded protocol instances, thereby establishing safety. As noted in the MLDR case study, these proofs were primarily completed by LLM agents with minimal human intervention.

\smallskip
\noindent\textbf{Benchmark coverage.}
The current 29-protocol suite covers classic distributed protocol models but may not represent the full diversity of real-world systems. Protocols with substantially different structures (such as more complex message-passing patterns, larger parameter domains, or deeply nested invariant clauses) may challenge \tool in ways not captured by the current suite.

\section{Related Work}\label{sec:relate}

\noindent\textbf{Automated invariant inference.}
Several tools have been developed to automatically infer inductive invariants for distributed protocols. DistAI~\cite{yao2021distai} takes a data-driven approach, enumerating candidate invariants from simulation traces and refining them with SMT-based counterexample feedback. I4~\cite{ma2019i4} infers invariants by model-checking finite protocol instances and generalizing the results to unbounded settings. SWISS~\cite{hance2021swiss} performs symmetry-aware enumerative synthesis over quantified invariant candidates. These tools operate in restricted first-order settings with decidable verification backends, limiting their applicability to expressive languages like TLA+. IC3PO~\cite{goel2021ic3po} extends IC3 with symmetry-aware quantified reasoning, exploiting indistinguishable domain sorts to lift finite-instance learned clauses to quantified invariants; however, its core method is specialized to protocols with strong structural regularity and does not naturally extend to invariants involving ordering, arithmetic, or non-symmetric relational structure. Its finite-to-unbounded convergence argument relies on additional unbounded checks rather than a fully closed theoretical guarantee.

Beyond enumerative approaches, Ivy~\cite{padon2016ivy} provides an interactive environment for safety verification and Feldman et al.~\cite{feldman2019phase} develop user-guided inference based on phase structures.
In more automated directions, Schultz et al.~\cite{schultz2024proofslicing} combine inductive proof slicing with grammar-based search, Zhang et al.~\cite{zhang2024kondo} introduce an invariant taxonomy separating Regular Invariants from Protocol Invariants, and Basilisk~\cite{zhang2025basilisk} automates proofs using provenance invariants in a Dafny-based workflow.

Endive~\cite{schultz2022kvy} operates directly on TLA+ specifications using a flat counterexample-to-induction (CTI) elimination loop, with predicates drawn from a manually curated pool. It demonstrated that CTI-guided invariant inference is practical for TLA+ protocols without restricting the model to decidable logic fragments. \tool shares Endive's emphasis on CTI-guided strengthening, but replaces flat CTI elimination with explicit IC3 frame management and replaces the manual predicate pool with LLM-generated blocking clauses. Compared with EPR-based, interactive, and Dafny-based approaches such as Basilisk, our method targets the full expressiveness of TLA+ while fitting directly into an existing TLA+/TLAPS workflow, avoiding translation or rewriting into another verification language.

IC3~\cite{bradley2011sat,bradley2012understanding} and its predecessor interpolation-based model checking~\cite{mcmillan2003interpolation} have been influential in hardware and protocol verification. Extensions to quantified and infinite-state reasoning~\cite{hoder2012generalized,cimatti2014ic3,cimatti2016ic3ia} have broadened IC3's applicability, while first-order quantified separators~\cite{koenig2020first} address quantified invariant inference in first-order settings. Our work adopts IC3's core control structure directly: frontier bad states are traced backward across shallower frames, learned clauses are admitted to specific frames, push attempts move clauses forward, and the search stops when adjacent frames coincide. The main difference from SAT-based PDR lies in clause discovery: \tool uses Apalache-generated bad states together with LLM-generated protocol-level blocking clauses instead of Boolean-cube generalization.

\smallskip
\noindent\textbf{TLA+ tools and verified systems.}
TLC~\cite{yu1999model} is the standard explicit-state model checker for TLA+, Apalache~\cite{konnov2019tla} provides symbolic model checking using SMT solvers, and TLAPS~\cite{chaudhuri2008tlaps} supports hierarchical proof development and proof checking---together forming a mature ecosystem with significant industrial adoption~\cite{newcombe2015aws}. Discovering inductive invariants, however, remains the primary bottleneck in the TLA+ verification workflow. Our tool attempts to complement this ecosystem by automating that step.

\smallskip
\noindent\textbf{LLMs for formal verification.}
Recent work has explored using LLMs for formal verification tasks. For instance, for loop invariant generation, Kamath et al.~\cite{kamath2024leveraging} prompt LLMs to generate loop invariants and candidate ranking functions
from natural-language descriptions of their definitions, and
then identifies valid candidates using Frama-C. LaM4Inv~\cite{wu2024llm} combines LLMs with bounded model checking in a query-filter-reassemble loop, and Chakraborty et al.~\cite{chakraborty2023ranking} propose contrastive ranking to prioritize correct invariants among LLM-generated candidates. Clause2Inv~\cite{cao2025clause2inv} proposes a generate-combine-check framework for loop invariant synthesis. Pirzada et al.~\cite{ibmc24} use LLMs to generate invariants to facilitate bounded model checking.

For broader program verification, Lemur~\cite{wu2024lemur} integrates LLMs into automated verification; Wen et al.~\cite{wen2024enchanting} combine LLMs with static analysis for specification synthesis. Quokka~\cite{wei2025quokka} presents a first-order framework for LLM-based invariant synthesis, using quantifier elimination to verify candidate invariants, and introduces the InvBench benchmark.

Relatively few LLM-based verification approaches operate directly on TLA+ specifications. (A notable exception is the recent work  by Zhou and Tripakis~\cite{zhou2025tlaplus}, which explores LLM-guided proof automation for TLA+.) One reason is that TLA+ distributed protocol models require quantified invariants over unbounded numbers of processes and complex protocol-level reasoning. \tool targets that setting. It performs inductive-invariant synthesis with an explicit IC3 controller and LLM-generated blocking clauses over TLA+ states. This enables end-to-end invariant synthesis without domain-specific predicate configuration. 
 
\section{Conclusion}\label{sec:concl}

We presented \tool, an IC3-based TLA+ tool that combines an explicit IC3 controller with LLM-generated blocking clauses. Across 29 distributed protocols including industrial-scale benchmarks such as MLDR and one complex Paxos variant that all four compared tools could not solve, \tool discovers candidate invariants for all 29 protocols, without manual template design. TLAPS further confirms that the discovered invariants are inductive for unbounded protocol instances.

The key insight behind these results is that LLMs need not produce complete invariants; rather they are most effective as \emph{hypothesis generators} within a structured IC3 architecture. The IC3 controller decomposes the global synthesis problem into focused per-state blocking tasks, validates every proposed clause through frame admission and forward propagation, and provides a completeness backstop when the LLM falls short. This division of labor, i.e., LLM for protocol-level conjecture and IC3 for proof discipline, underpins the robustness of our approach: even weaker (or differently trained) models can solve a substantial fraction of benchmarks because the surrounding verification structure compensates for imprecise clause proposals.

More broadly, \tool suggests that formal verification is not necessarily an exclusive domain of verification experts. By embedding LLM reasoning inside a principled proof architecture, \tool makes inductive-invariant discovery accessible to protocol designers who can write a TLA+ specification but lack the expertise to craft strengthening invariants manually. We view this as a step toward making formal methods a routine of distributed systems development.

\subsubsection*{Future work.} We plan to explore the following directions.
\smallskip 
\noindent\emph{Scaling.} We plan to apply \tool to larger, more complex protocols from real engineering settings with richer state spaces, and to further optimize the IC3 controller and the LLM-guided clause-learning strategy.  

\smallskip
\noindent\emph{Agent-based workflow.}
A natural next step is to equip \tool with LLM agents that invoke verification skills, tools and scripts. Our attempt in this direction suffered from premature stopping and high latency due to repeated agent tool and script calls, making it less efficient than the current specialized pipeline and not yet capable of effective unattended inductive-invariant derivation. However, we believe that a more carefully designed agent workflow can overcome these limitations and further enhance the system's ability to derive stronger inductive invariants without manual effort.

\smallskip 
\noindent\emph{Automating TLAPS discharge.}
Recent work on LLM-guided TLA+ proof automation~\cite{zhou2025tlaplus} and agent-based proof generation~\cite{yang2025verusage} shows promise for discharging proof obligations once the key lemmas are available; however, inferring the inductive invariants that those lemmas require remains beyond the reach of current proof agents~\cite{yang2025verusage}. \tool is well positioned to close this gap: it can organize the validated invariants it discovers into candidate lemmas and feed them to TLAPS~\cite{chaudhuri2008tlaps}, so that proof agents can focus on identifying the remaining proof goals and then generating, refining and repairing proof scripts automatically.

\clearpage

\bibliographystyle{ACM-Reference-Format}
\bibliography{ref}

\end{document}